\documentclass[preprint]{aastex}
\def\sss{\scriptscriptstyle}
\def\^#1{^{\sss #1}}
\def\_#1{_{\sss #1}}
\def\beq{\begin{equation}}
\def\eeqno#1{\label{#1}\end{equation}}

\def\rarrow{\rightarrow }

\def\dleft{\rlap{{\it D}}\raise 8pt\hbox{$\scriptscriptstyle\Leftarrow$}}
\def\dright{\rlap{{\it
D}}\raise 8pt\hbox{$\scriptscriptstyle\Rightarrow$}}

\def\cmss{{\rm cm~s^{-2}}}

\def\pc{{\rm pc}}

\def\msun{M\_{\odot}}
\def\az{a\_{0}}

\def\l0{\ell\_{0}}

\def\l{\lambda}

\def\f{\phi}

\def\r{\rho}
\def\rp{\rho_p}

\def\m{\mu}

\def\U{\mathcal{U}}

\def\xlimin{{x\rarrow\infty \atop{\raise 1pt\hbox to 30pt{\rightarrowfill}}}}
\def\limlim#1#2{{#1\rarrow #2 \atop{\raise 1pt\hbox to 30pt{\rightarrowfill}}}}

\def\ve{{\bf e}}

\def\vg{{\bf g}}
\def\vgN{{\bf g}\_N}

\def\ve{{\bf e}}

\def\S{\Sigma}

\def\grad{\vec\nabla}
\def\div{\vec \nabla\cdot}
\def\gf{\grad\phi}

\def\SM{\Sigma\_M}
\def\Sc{\Sigma_c}

\def\c{\gamma}

\def\m{\mu}

\def\r{\rho}
\def\l{\lambda}

\begin{document}

\title{The central surface density of ``dark halos'' predicted by MOND}
\author{Mordehai Milgrom }
\affil{ The Weizmann Institute Center for Astrophysics, Rehovot,
76100, Israel}

\begin{abstract}
Prompted by the recent claim, by Donato \& al., of a quasi-universal
central surface density of galaxy dark matter halos, I look at what
MOND has to say on the subject. MOND, indeed, predicts a
quasi-universal value of  this quantity for objects of all masses
and of any internal structure, provided they are mostly in the
Newtonian regime; i.e., that their mean acceleration is at or above
$\az$.  The predicted value is $\c\SM$, with $\SM\equiv \az/2\pi G=
138(\az/1.2\times 10^{-8}\cmss)\msun\pc^{-2}$, and $\c$  a constant
of order 1 that depends only on the form of the MOND interpolating
function. For the nominal value of $\az$,
$\log(\SM/\msun\pc^{-2})=2.14$, which is consistent with that found
by Doanato \& al. of $2.15\pm0.2$.

MOND predicts, on the other hand, that this quasi-universal value is
not shared by objects with much lower mean accelerations. It permits
halo central surface densities that are arbitrarily small, if the
mean acceleration inside the object is small enough. However, for
such low-surface-density objects, MOND predicts a halo surface
density that scales as the square root of the baryonic one, and so
the range of the former is much compressed relative to the latter.
This explains, in part, the finding of Donato \& al. that the
universal value applies to low acceleration systems as well. Looking
at literature results for a number of the lowest surface-density
disk galaxies with rotation-curve analysis, I find that, indeed,
their halo surface densities are systematically lower then the above
``universal'' value.

The prediction of $\SM$ as an upper limit, and accumulation value,
of halo central surface densities, pertains, unlike most other MOND
predictions, to a pure ``halo'' property, not to a relation between
baryonic and ``dark matter'' properties.

\end{abstract}


\section{introduction}
Donato \& al. (2009) have recently looked at the distribution of the
central surface densities, $\Sc$, of the dark matter halos
(hereafter CHSD) of galaxies of different types. They find that the
distribution is rather narrow, with a central value
$\Sc=10^{2.15\pm0.2}\msun\pc^{-2}$. This finding agrees with
previous studies, in particular with that of Milgrom \& Sanders
2005, who dealt with the relevance to MOND, and with others (see
references in Donato \& al. 2009). $\Sc$ is defined by Donato \& al.
as the product of the central halo density, $\r_0$, and the core
radius, $r_0$, both derived by fitting halo-plus-baryons models to
various observations, such as rotation curves, weak lensing results,
or velocity dispersion data. In deducing $\r_0$ and $r_0$, the halo
is sometimes assumed to have a density distribution of the cored
isothermal form; Donato \& al. assumed a spherical Burkert profile.
\par
A surface density of special role, $\Sc$, translates into an
acceleration of a special role $\Sc G$, and this immediately evokes
MOND. One is thus naturally led to consider whether such a special
value for the CHSD is predicted by MOND.
\par
Brada \& Milgrom (1999) showed that MOND predicts an absolute
acceleration maximum, of order $\az$, that any phantom halo can
produce, anywhere in an object. Milgrom \& Sanders (2005), in a
precursor to Donato \& al. (2009), tested this MOND prediction by
plotting, for a sample of 17 Ursa Major galaxies, the deduced $\r_0$
and $r_0$ against each other. (These were deduced for a cored
isothermal sphere model, not a Burkert one, with a variety of
assumptions on the stellar $M/L$ values: maximum disc, population
synthesis values, best MOND fits to rotation curves, etc.) They
found, for their sample, that these parameters lie near a line of
constant $\Sc=10^2\msun\pc^{-2}$ (their Fig.4), in agreement with
the value Donato \& al. find. This was interpreted by Milgrom \&
Sanders (2005) as indicating a maximum halo acceleration as
suggested by Brada \& Milgrom (1999), because the sample used was
devoid of truly low-surface brightness galaxies, for which ``halo''
accelerations are supposedly lower.
\par
Here I will show, as a new result, that MOND does indeed predict a
quasi-universal value for the CHSD of the imaginary, or phantom,
dark matter (DM), but only for baryonic systems that are, by and
large, in the Newtonian regime, having mean internal accelerations
of order $\az$ or larger. In contradistinction, MOND predicts that,
in principle, we can have galaxies with arbitrarily small values of
$\Sc$, if the baryonic surface density is low enough. However, the
predicted CHSD scales as the square root of the baryonic surface
density, and so will have a rather contracted span in a given
sample.
\par
Of course, each of the objects in the sample studied can, and
should, be used to subject MOND to a detailed, individual test.
Inasmuch as MOND passes theses tests, as it seems to do quite well,
we can deduce that there is an acceptable halo model whose analog of
$\Sc$ agrees with the MOND prediction. If other halo models do not
agree with the MOND prediction, it only shows that there is a range
of acceptable halo parameters, within the uncertainties in the model
parameters or assumptions (assumed density law for the halo, stellar
$M/L$ values, etc.).
\par
Individual tests are, collectively, more decisive than tests of
general rules, which they subsume. Nevertheless, deducing and
testing such general rules, such as the mass-rotational-speed
relation (aka the baryonic Tully-Fisher relation), or the MOND
prediction underlying the Faber-Jackson relation, have obvious
merits of their own, as they focus attention on certain unifying
principles. In this light it is important to consider the prediction
of a quasi-universal CHSD in itself.
\par
In section \ref{universal}, I explain how the quasi-universal CHSD
arises in MOND, for high-acceleration systems. In section \ref{lsb},
I treat systems with low surface density; in particular, I show from
results in the literature that disk galaxies with the lowest surface
densities analyzed to date, do have $\Sc$ values that fall
systematically below the quasi-universal value.  The discussion
section \ref{discussion} deals with the special significance of the
prediction at hand, in comparison with other MOND predictions.

\section{The emergence of a quasi-universal ``halo'' central
surface density in high acceleration systems} \label{universal} I
shall be using the formulation of MOND as modified gravity put forth
by Bekenstein \& Milgrom (1984). In this theory the MOND
gravitational potential, $\f$, is determined by a nonlinear
generalization of the Poisson equation
 \beq \div[\m(|\gf|/\az)\gf]=4\pi G\r,  \eeqno{poisson}
 $\r$ being the true (``baryonic'')  matter density.
Here $\m(x)$ is the interpolating function characterizing the
theory, and $\az$ is the MOND acceleration constant, known from
various analyses to be $\az\approx 1.2\times 10^{-8}\cmss$ (see,
e.g., Stark, McGaugh, \& Swaters 2009 who find that gas dominated
galaxies satisfy the mass-asymptotic-rotational-velocity relation
predicted by MOND, $M=\az^{-1}G^{-1}V^4_{\infty}$, with this value
of $\az$). Similar results will follow from the pristine, algebraic
formulation of MOND (Milgrom 1983). Also, if the halo properties are
derived from rotation-curve analysis, the same results will follow
in modified inertia theories, since these theories predict the
algebraic relation between the Newtonian and MOND accelerations for
circular orbits. We do not know exactly what these modified inertia
theories say about gravitational lensing, but we expect similar
results from this as well. Regarding lensing, the existing
relativistic extension of the modified-Poisson theory, TeVeS (see
Bekenstein 2006 and Skordis 2009 for reviews), says that we can use
the halo as deduced from the modified Poisson theory to derive
lensing in the standard way, at least when we can assume approximate
spherical symmetry. Weak-lensing halo properties can thus be
compared directly with the predictions of this theory.

\par
When interpreted by a Newtonist, the departure predicted by MOND,
and encapsulated in the difference between the MOND acceleration
field $\gf$ and the Newtonian one, is explained by the presence of
``dark matter'', or ``phantom matter'' whose density is (Milgrom
1986)
 \beq \rp={1\over 4\pi G}\Delta\f-\r. \eeqno{phantom}
\par
Using the field equation (\ref{poisson}) we can write
 \beq \rp=-{1\over 4\pi
G\az}(\m'/\m)\grad|\gf|\cdot\gf+(\m^{-1}-1)\r, \eeqno{phantomu}
 which can be cast in another form
 \beq \rp=-{\az\over 4\pi G}\ve\cdot\grad\U(|\gf|/\az)+(\m^{-1}-1)\r,
\eeqno{phantoma} where $\U(x)=\int L(x)dx$, with $L=x\m'/\m$  the
logarithmic derivative of $\m$, and $\ve$ is a unit vector in the
direction of $\gf$. This form is particularly useful for calculating
column densities of $\rp$ along field lines, as we want to do here.
This relation is exact. Expression (\ref{phantoma}), with $\gf$
replaced by $-\vg$, holds exactly in the more primitive, algebraic
formulation, whereby the MOND acceleration $\vg$ is given by
$\m(|\vg|/\az)\vg=\vgN$; $\vgN$ being the Newtonian acceleration;
$\vg$ is not generally derivable from a potential.
\par
Consider first an arbitrary point mass, and integrate expression
(\ref{phantoma}) along a line through the point mass. This gives the
central surface density of the phantom matter halo surrounding the
mass, $\S(0)$. Inside the mass $\m\approx 1$ so the second term does
not contribute. The integral is performed in two segments: from
$-\infty$ to the point mass (where $\ve$ is opposite the direction
of integration) and from the other side of the point mass to
$\infty$. The two combined give
 \beq \S(0)=\int_{-\infty}^{\infty}\rp
 dz=\SM[\U(\infty)-\U(0)]=\SM\int_0^{\infty}L(x)dx\equiv
  \l\SM,  \eeqno{sigma}
where,
 \beq \SM\equiv {\az\over 2\pi G} \eeqno{sigmamond}
is the relevant surface density proxy for $\az$ in the present
context. In the deep MOND regime ($x\ll 1$) $L(x)\approx 1$ , and
far outside the MOND regime $L(x)\approx 0$; so $\l$ is of order 1,
and depends only on the interpolating function $\m(x)$.
\par
 I am dealing all along with central column density
$\S(0)=2\int_0^{\infty}\r dr$ of the MOND phantom halo. For a
Burkert halo this column density is related to the quantity $\Sc$,
used by Donato \& al., by $\S(0)=(\pi/2)\Sc$. So, translating the
column density to the MOND analog of $\Sc$, call it $\Sc^*$,
 \beq \Sc^*=(2\l/\pi)\SM\equiv \c\SM.  \eeqno{kulya}
We have
 \beq \SM= 138(\az/1.2\times
10^{-8}\cmss)\msun\pc^{-2}, \eeqno{sigmam}
 or, for the nominal value
of $\az$, $\log(\SM/\msun\pc^{-2})=2.14$, compared with the value
$\log(\Sc/\msun\pc^{-2})=2.15\pm0.2$ found by Donato et
al.\footnote{The predicted MOND ``halo'' of an isolated system is
not well described by a Burkert profile: The MOND ``halo'' density
behaves asymptotically as $r^{-2}$, not $r^{-3}$, and it is expected
to have a depression around the center not a decreasing density
profile everywhere. Nevertheless, these differences are expected to
produce only differences by a factor of order 1 in the resulting
$\Sc$. The very near equality of $\SM$ and the central value found
by Donato \& al. is thus somewhat fortuitous.}.
\par
  For the limiting form of
$\m(x)$--with $\m(x)=x$, for $x\le 1$, and $\m(x)=1$, for $x>1$--we
have $\l=1$, and $\c=2/\pi$. For $\m(x)=x(1+x^2)^{-1/2}$, we have
$\l=\pi/2$, and $\c=1$. Values of $\l$ for other forms of $\m$ can
be read off Fig. 3 of Milgrom \& Sanders (2008) (where they were
deduced numerically, and appear for other purposes). One sees that
$1\lesssim\l\lesssim 3$, and so $0.7\lesssim\c\lesssim 2$ for the
range of $\m$ forms studied there\footnote{The coefficient $\l$
diverges if $1-\m(x)$ behaves at large $x$ as $x^{-1}$ or slower.
The divergence does not occur in the MOND regime, but comes from the
Newtonian regime very near the point mass. Such a behavior of $\m$
is, however excluded strongly from solar system constraints, and I
preclude it.}.
\par
Equations (\ref{sigma})-(\ref{kulya}) are our basic result, around
which all else in the paper revolves. They tell us that for the
simple case of a point mass a universal value of $\Sc$ is indeed
predicted by MOND; its value is $\approx\SM$, which agrees very well
with the value found by Donato \& al..
\par
Consider now an extended mass, $M$. If the mass is well contained
within its MOND transition radius, $R\_M=(MG/\az)^{1/2}$, namely if
the Newtonian accelerations, and hence the MOND accelerations, are
high everywhere within the mass, then the procedure we followed for
a point mass applied approximately, and we get again $\Sc^*\approx
\c\SM$.
\par
 Here I have to pause, and comment on a subtlety in the use of
eq.(\ref{phantoma}), and in interpreting the results thereof. This I
demonstrate with two examples. First consider a mass of finite
extent whose density does not increase towards its center as
$r^{-1}$ or faster. In this case, the Newtonian acceleration, and so
also the MOND acceleration, goes to zero at the center, even if
these accelerations are much higher than $\az$ in most of the bulk.
In other words, there are two MOND regimes: one within some small
sphere of radius $r_1$ around the center, and another beyond the
MOND transition radius, $R\_M=(MG/\az)^{1/2}$. The small $r$ region
contributes to $\S(0)$ through the first term in
eq.(\ref{phantoma}), an amount $-\SM\int_0^{X_0}L(x)dx$, where $X_0$
is the maximum (MOND) acceleration in units of $\az$. This
contribution is $\approx -\l\SM$ for $X_0\gg 1$. The outer region
contributes a positive quantity of the same magnitude. In addition,
the inner region contributes through the second term in
eq.(\ref{phantoma}), and its total contribution is positive (the
phantom density is always positive in the spherical case).  The
inner region of phantom mass, even if it contributes to $\S(0)$, has
only little mass, is dynamically unimportant, at large, and should
not be included when comparing with results for global halo
parameters. I shall thus ignore it, and  take  $\l\approx
\int_0^{X_0}L(x)dx$.
 When the baryonic surface density is
low, the central, low-acceleration region is expanded and
encompasses the whole mass. The contribution of the first term in
eq.(\ref{phantoma}) then can, indeed, be taken to vanish, and the
contribution to $\S(0)$ comes from the second term.
\par
In another example,  consider two arbitrary point masses along the
line of sight. Integrating the phantom density in
eq.(\ref{phantoma}) along the line of sight now gives
$\S(0)=2\l\SM$. (We now have to integrate over four segments over
which $\ve$ changes sign: from $-\infty$ to the first mass, from
there to the zero-field point somewhere between the masses, from
there to the second mass, and from there to infinity). This value is
exact and independent of the distance between the masses. How is
this consistent then with  our deduction that $\S(0)\approx \l\SM$
for all systems well within their transition radius? When the two
masses are well separated, by more then their joint transition
radius, there is an extended halo surrounding each of the masses,
each halo with its own $\S(0)\approx \l\SM$, and the two column
densities add up. When the two masses are near each other, well
within their joint $R\_M$, there will be a common halo of phantom
matter residing roughly beyond $R\_M$, and this indeed has
$\S(0)\approx \l\SM$ [arising from integrating eq(\ref{phantoma}) in
the outer two segments]. In addition, there is a small region around
the point of zero field between the two masses, which contributes
the same amount to the central column density, but which contains
little mass, is dynamically unimportant in the present context, and
should be eliminated from the result that is to be compared with the
observations.
\par
Keeping these caveats in mind, the reasoning leading to
eq.(\ref{sigma}) can be applied not only to spherical systems. For
example, for a disk galaxy with a high central surface (baryonic)
density, $\S_b(0)\gg\SM$, we can use this equation to calculate the
column density either along the symmetry axis, or along a diameter
in the plane of the disc (in both cases the field is always parallel
or antiparallel to the line of integration). If we ignore the small
region of phantom matter near the very center (or if we add a small
matter cusp that prevents the acceleration from vanishing at the
center) we again get $\S(0)=\l\SM$.

\par
Take now, more generally the extent of our mass to be $R$, and its
mean density $\r$, and define $\S_b=\r R$, the baryonic equivalent
of $\Sc$. The second term in eq.(\ref{phantoma}) can be estimated to
contribute to $\S(0)$
 \beq \approx 2\r R[\m^{-1}(g/\az)-1], \eeqno{utrewq}
where $g$ is the MOND mean acceleration inside the mass, and is
given by
 \beq (g/ \az)\m(g/\az)\approx{4\pi\over 3}{\r R G\over \az}={2\over 3}{\S_b\over
 \SM}.  \eeqno{byrew}

 The first term in eq.(\ref{phantoma}) is taken to contribute $\approx\SM\int_0^{X_0}L(x)dx$, where $X_0=g/\az$.
Thus, we can write
  \beq \S_c^*=(2/\pi)\S(0)\approx \SM \{(6/\pi)X_0[1-\m(X_0)]+\int_0^{X_0}L(x)dx\}. \eeqno{guplo}
For $X_0\gg 1$ this gives $\Sc^*\approx\c\SM$, again.

\section{Low surface density systems}
\label{lsb} MOND does permit arbitrarily low values of $\Sc$ for
phantom halos in low acceleration systems. When $X_0\ll 1$, namely,
when the maximum (MOND) acceleration in the system is much smaller
than $\az$, we get from  eq.(\ref{guplo}), to lowest order in $X_0$,
 \beq \Sc^*\approx(6/\pi+1)\SM X_0\approx
 2.4\left({\S_b\over \SM}\right)^{1/2}\SM. \eeqno{gutsa}
 Such low acceleration systems are characterized by low baryonic
 surface densities $\S_b/ \SM\ll 1$. Note, however, that the departure  from the universal $\Sc^*$
sets in at rather low baryonic surface densities, since $\Sc^*/\SM$
scales as the square root of $\S_b/ \SM$. The lowest acceleration
disc galaxies studied to date have $X_0$ values only down to
0.1-0.2; and we see from eq.(\ref{gutsa}) that
 even for values of $X_0$ as low as 1/5 we get
$\Sc^*\approx 0.6\SM$. Clearly, however, MOND does predicts that,
for extremely low baryonic surface density galaxies, the CHSD falls
increasingly below the quasi-universal value.
\par
To superficially check this expectation, I looked (rather randomly)
in the literature for derived halo parameters for the lowest
acceleration disk galaxies with rotation-curve analysis. Three such
galaxies were analyzed in light of MOND by Milgrom \& Sanders
(2007), showing rather satisfactory agreement. These were also
analyzed earlier in terms of cored isothermal halos: For KK98 250
and KK98 251, I find in Begum \& Chengalur (2005) best-fit
parameters that give $\Sc= 56$, and $66 \msun/\pc^2$, respectively.
For NGC 3741, Begum \& al. (2005) find parameters that yield
$\Sc=56\msun/\pc^2$. All three values fall substantially below the
nominal quasi-universal value of $\Sc=140\msun/\pc^2$, and are
consistent, within the uncertainties, with our rough estimate
(\ref{gutsa}), having $X_0$ values of between 0.1 and 0.3. The first
two galaxies were not included in the Donato \& al. analysis; but
NGC 3741 was included, based on the analysis of  Gentile \& al.
(2007) (assuming a Burkert, not a cored isothermal halo), whose
results give $\Sc=74$. This value is higher than the result of Begum
\& al. (2005) (though consistent within the uncertainties), but
still only about half the quasi-universal value.
\par
Another low acceleration galaxy that is worth analyzing in detail
(and is not included in the Donato \& al. sample) is the dwarf
Andromeda IV. Its rotation curve is given in Chengalur \& al.
(2007). To my knowledge, its photometry and HI distribution are not
yet available publicly for rotation curve analysis. However,
according to Chengalur et \&. (2007) it is heavily dominated by gas
with $M\_{gas}/L\approx 18$, and it shows a very strong mass
discrepancy with $M\_{dyn}/M\_{gas}\approx 14$ at the last measured
point. In deriving a cored isothermal halo parameters we can thus
approximately ignore the baryons and fit the rotation curve with the
halo alone. Doing this, I find, tentatively, $\Sc\sim
45\msun/\pc^2$, about three times lower than $\SM$. Since in this
case $X_0\sim 0.1-0.15$, this is also in agreement with the estimate
of eq.(\ref{gutsa}).
\par
Why then do Donato \& al. suggest that the quasi-universal value of
$\Sc$ applies to all galaxies, including the very low-acceleration
ones? This is based mostly on the analysis of dwarf spheroidal
satellites of the Galaxy. Their analysis includes only one well
studied low-acceleration disk, the above mentioned, NGC 3741--for
which, as we saw, the actual $\Sc$ could be lower--and one somewhat
higher acceleration galaxy, DDO 47. As regards the dwarf spheroidal
Milky-Way satellites, MOND would indeed predict lower values of
$\Sc$  than adopted by Donato \& al.. However, as Donato \& al.
emphasize themselves, the analysis of these systems is beset by
uncertainties in the model assumptions (e.g., assumptions on orbital
anisotropies), leading to non-unique results. Angus (2008) has
analyzed these dwarf spheroidals in MOND, and found that, with two
exceptions perhaps, they can be well explained by MOND, assuming
appropriate orbit anisotropy distributions. This would mean, as I
stressed above, that there are acceptable ``halo'' models that are
consistent with the predictions of MOND. The disparate values
adopted by Donato \& al. only demonstrate the non-uniqueness of the
halo-parameter determination.

\section{Discussion}
\label{discussion} I have shown that the acceleration constant of
MOND $\az$ defines a special surface density parameter $\SM=\az/2\pi
G$. This serves as a quasi-universal central surface density of
phantom halos around objects of all masses and structures, provided
they are themselves in the Newtonian regime (i.e., with bulk
accelerations of order $\az$ or higher).
\par
This is a particularly interesting prediction of MOND, because most
of the other salient MOND predictions relate properties of the true
matter (baryons) to those of the putative dark matter halo. This is
the case for the mass-velocity (baryonic Tully-Fisher) relation, the
Faber-Jackson relation, the transition from baryon dominance to DM
dominance at a fixed acceleration, the full prediction of rotation
curves, the necessity of a disk component of DM, in disk galaxies,
in addition to a spheroidal halo, etc. (see Milgrom 2008 for a more
detailed list, and explanations). Here, however, we have a
prediction that speaks of a property of halos themselves, without
regard to the true mass that engenders them, apart from the
requirement that the baryons be well concentrated. $\SM$ may also be
viewed as an upper limit, and accumulation value, for ``halo''
central surface densities, irrespective of baryonic properties. It
is clear then, that the $\az$ that appears in this prediction need
have nothing to do with the $\az$ that appears in other relations,
in the framework of the DM doctrine. We could have a sample of halos
all satisfying the present prediction, and add to them baryons
arbitrarily, so as not to satisfy, e.g., the baryonic-mass-velocity
relation $MG\az=V^4$, which also revolves around some acceleration
constant. The fact that the $\az$ emerging from the phenomenology
here is the same as that appearing in the other phenomenological
relation should be viewed as another triumph of MOND.
\par
There are two other MOND predictions that speak of properties of the
halo alone. The first is that the density profile of the ``halo'' of
any isolated object behaves asymptotically as $r^{-2}$ (asymptotic
flatness of rotation curves). The other such prediction is the
maximally allowed acceleration (of order $\az$) that a halo can
produce (Brada \& Milgrom 1999). This is simply a reflection of the
MOND tenet that the phantom mass cannot be present where
accelerations are higher than roughly $\az$. The prediction I
discuss here can be understood, qualitatively, as a result of the
above two: On the asymptotic, $r^{-2}$, tail of the ``halo'', the
acceleration it produces is $g_h\approx 4\pi G\r r$. Going inward,
the maximum-acceleration prediction tells us that this behavior can
continue only down to a radius where $g_h\sim \az$. Below this
radius the halo density profile must become shallow and produce a
core. This gives $\r_0 r_0\sim\az/4\pi G$. It is this that underlies
our more quantitative result here.
\par
However, there is nothing in MOND to forbid the halo density profile
from becoming shallow within a radius much larger then that where
$g_h\sim\az$. This can happen at arbitrarily large radii, producing
arbitrarily small values of $\Sc^*$, as indeed our detailed analysis
shows.

\section*{Acknowledgements}

 This research was supported by a
center of excellence grant from the Israel Science Foundation.

\end{document}